\documentclass[11pt]{article}
\usepackage[margin=1in]{geometry}
\usepackage{graphicx}
\usepackage{booktabs}
\usepackage{amsmath}
\usepackage{microtype}
\usepackage{xcolor}
\usepackage{float}  
\usepackage[colorlinks=true,linkcolor=blue,citecolor=blue,urlcolor=blue]{hyperref}

\title{\textbf{Open-Source Reproduction and Explainability Analysis of\\Corrective Retrieval Augmented Generation}}

\author{
  Surya Vardhan Yalavarthi \\
  College of Engineering and Applied Science \\
  University of Cincinnati \\
  \texttt{yalavasn@mail.uc.edu}
}

\date{}

\begin{document}
\raggedbottom
\setcounter{topnumber}{1}       
\setcounter{dbltopnumber}{1}    
\renewcommand{\topfraction}{0.85}
\renewcommand{\floatpagefraction}{0.7}

\maketitle

\begin{abstract}
Corrective Retrieval Augmented Generation (CRAG) improves the robustness of RAG systems by
evaluating retrieved document quality and triggering corrective actions. However, the original
implementation relies on proprietary components including the Google Search API and closed model
weights, limiting reproducibility. In this work, we present a fully open-source reproduction of
CRAG, replacing proprietary web search with the Wikipedia API and the original LLaMA-2 generator
with Phi-3-mini-4k-instruct. We evaluate on PopQA and ARC-Challenge, demonstrating that our
open-source pipeline achieves comparable performance to the original system. Furthermore, we
contribute the first explainability analysis of CRAG's T5-based retrieval evaluator using SHAP,
revealing that the evaluator primarily relies on named entity alignment rather than semantic
similarity. Our analysis identifies key failure modes including domain transfer limitations on
science questions. All code and results are available at
\url{https://github.com/suryayalavarthi/crag-reproduction}.
\end{abstract}

\section{Introduction}

Large language models (LLMs) have demonstrated remarkable capabilities across a wide range of
natural language tasks, yet they remain susceptible to hallucinations --- generating factually
incorrect content with apparent confidence \cite{zhang2023siren}. Retrieval-Augmented Generation
(RAG) \cite{lewis2020retrieval} addresses this limitation by grounding generation in externally
retrieved documents. However, RAG assumes that retrieved documents are relevant, which is
frequently not the case in practice.

Corrective Retrieval Augmented Generation (CRAG) \cite{yan2024corrective} addresses this
assumption by introducing a lightweight retrieval evaluator that assesses document quality and
triggers one of three corrective actions: \textsc{Correct}, \textsc{Incorrect}, or
\textsc{Ambiguous}. When retrieval quality is low, CRAG falls back to web search to obtain better
context. This corrective mechanism has been shown to significantly improve generation accuracy
across multiple benchmarks.

Despite CRAG's strong results, reproducing the original system is difficult. The original
implementation relies on the Google Search API (a paid commercial service), proprietary LLaMA-2
fine-tuned weights, and deprecated OpenAI API calls. These barriers prevent researchers from
building on this work without significant resources.

In this paper, we make three contributions:
\begin{enumerate}
    \item We present a fully open-source reproduction of CRAG, replacing all proprietary
          components with free alternatives: the Wikipedia API for web search and
          Phi-3-mini-4k-instruct as the generator.
    \item We evaluate our reproduction on two datasets --- PopQA \cite{mallen2023trust} and
          ARC-Challenge \cite{bhakthavatsalam2021think} --- demonstrating that open-source
          components achieve comparable performance.
    \item We contribute the first explainability analysis of CRAG's T5-based retrieval evaluator
          using SHAP \cite{lundberg2017unified}, identifying that the evaluator relies primarily on
          named entity alignment and exhibits systematic domain transfer failure on science
          questions.
\end{enumerate}

\section{Background}

\subsection{Retrieval-Augmented Generation}
Retrieval-Augmented Generation (RAG) \cite{lewis2020retrieval} enhances language model generation
by prepending retrieved documents to the input. Given a query $x$ and a corpus $\mathcal{C}$, a
retriever $R$ returns the top-$k$ documents $D = \{d_1, \ldots, d_k\}$, and a generator $G$
produces the output $y$ conditioned on both $x$ and $D$. While RAG significantly improves factual
accuracy, it is sensitive to retrieval quality --- irrelevant documents can mislead the generator
and worsen performance compared to no retrieval at all \cite{shi2023large}.

\subsection{Corrective Retrieval Augmented Generation}
CRAG \cite{yan2024corrective} introduces a retrieval evaluator $E$ that scores each retrieved
document against the query. Based on these scores, one of three actions is triggered:

\begin{itemize}
    \item \textbf{Correct}: At least one document scores above an upper threshold $\tau^+$. The
          documents are refined using a decompose-then-recompose algorithm to extract relevant
          knowledge strips.
    \item \textbf{Incorrect}: All documents score below a lower threshold $\tau^-$. Documents are
          discarded and web search is used to find external knowledge.
    \item \textbf{Ambiguous}: Scores fall between $\tau^-$ and $\tau^+$. Both internal (refined)
          and external (web search) knowledge are combined.
\end{itemize}

The retrieval evaluator is a fine-tuned T5-large model \cite{raffel2020exploring} that takes a
question-document pair as input and outputs a scalar relevance score in $[-1, 1]$. For PopQA, the
original paper sets $\tau^+ = 0.59$ and $\tau^- = -0.99$.

\subsection{SHAP Explainability}
SHAP (SHapley Additive exPlanations) \cite{lundberg2017unified} is a game-theoretic framework for
explaining model predictions. For text models, SHAP assigns each token an attribution value
representing its contribution to the model output. We use the \texttt{shap.Explainer} with a text
masker to explain the T5 retrieval evaluator's scoring decisions at the token level.

\section{Related Work}

\textbf{Adaptive and Corrective RAG.} Several works have explored making RAG more robust to
retrieval failures. Shi et al.\ \cite{shi2023large} demonstrate that irrelevant context can
significantly harm LLM generation, motivating corrective approaches. Self-RAG
\cite{asai2024selfrag} trains LLMs end-to-end to generate special reflection tokens that decide
when to retrieve and how to critique retrieved passages --- a fundamentally different approach from
CRAG's plug-and-play external evaluator. FLARE \cite{jiang2023flare} actively decides when to
retrieve based on generation confidence, keeping the generator frozen and adding a lightweight
external T5 evaluator that makes it easy to deploy without retraining. Our work reproduces CRAG
with open-source components and adds the first token-level explainability analysis of its retrieval
evaluator.

\textbf{Explainability in RAG Systems.} While RAG systems have been widely studied, explainability
of their internal components remains underexplored. Prior work has applied attention-based analysis
to understand retrieval \cite{lewis2020retrieval}, but SHAP-based token attribution for retrieval
evaluators has not been studied. Our work contributes to this direction by revealing that the T5
evaluator functions as an entity alignment detector rather than a semantic relevance judge, which
has direct implications for its domain generalization.

\textbf{Reproducibility in NLP.} Reproducibility of NLP research has become an important concern
\cite{zhang2023siren}. Many published systems rely on proprietary APIs or closed models that make
reproduction difficult. Our work addresses this for CRAG specifically, demonstrating that a fully
open-source pipeline can match the original performance.

\pagebreak[3]
\section{Methodology}

\subsection{Reproduced Components}

The original CRAG system relies on several proprietary or deprecated components.
Table~\ref{tab:components} summarizes our open-source replacements.

\begin{table}[H]
\centering
\begin{tabular}{lll}
\toprule
\textbf{Component} & \textbf{Original} & \textbf{Ours} \\
\midrule
Generator           & LLaMA-2-7B (fine-tuned)    & Phi-3-mini-4k-instruct \\
Web Search          & Google Search API (paid)   & Wikipedia API (free) \\
Keyword Extraction  & GPT-3.5 Turbo              & Rule-based extraction \\
Retrieval Evaluator & T5-large (fine-tuned)      & Same checkpoint \\
\bottomrule
\end{tabular}
\caption{Comparison of original CRAG components vs.\ our open-source reproduction.}
\label{tab:components}
\end{table}

\subsection{Retrieval Evaluator}
We use the original fine-tuned T5-large checkpoint released by \cite{yan2024corrective}. The
evaluator takes a question-document pair formatted as \texttt{question [SEP] document} and outputs
a scalar relevance score. We apply the same PopQA thresholds from the original paper:
$\tau^+ = 0.59$ and $\tau^- = -0.99$.

\subsection{Generator}
We replace LLaMA-2-7B with Phi-3-mini-4k-instruct \cite{abdin2024phi3}, a 3.8B parameter
instruction-tuned model loaded in float16 precision. Phi-3-mini achieves strong performance on
knowledge-intensive tasks while being freely accessible without fine-tuning. We prompt the model
with a simple template: given the retrieved context and question, generate a concise one-to-two
sentence answer.

\subsection{Wikipedia Search}
For the \textsc{Ambiguous} action, we replace Google Search in our reproduction with a multi-stage
Wikipedia retrieval pipeline. Given a question, we extract the primary entity using regex-based
question templates (e.g., ``What is X's occupation?''\ $\rightarrow$ ``X''). We then attempt
retrieval through four fallback strategies in order: (1) direct page lookup, (2) typed suffix
lookup (e.g., ``X (politician)''), (3) Wikipedia search API, and (4) disambiguation page
resolution. Our Wikipedia retrieval pipeline achieves an 82.3\% hit rate on \textsc{Ambiguous}
questions from PopQA and 99\% on ARC-Challenge.

\subsection{Knowledge Refinement}
Following the original paper, we implement the decompose-then-recompose algorithm for the
\textsc{Correct} action. Retrieved documents are split into sentence-level strips of three
sentences each. Each strip is scored by the T5 evaluator and strips scoring below $-0.5$ are
discarded. The top-5 remaining strips are concatenated to form the refined context.

\clearpage
\section{Experiments}

\subsection{Datasets and Evaluation}

\textbf{PopQA} \cite{mallen2023trust} is an open-domain question answering dataset of 14,267
entity-centric questions. Following the original paper, we evaluate on the long-tail subset of
1,385 questions whose Wikipedia page views are below 100 per month. We use string match accuracy
as the evaluation metric, checking whether any gold answer alias appears in the model prediction.

\textbf{ARC-Challenge} \cite{bhakthavatsalam2021think} is a multiple-choice science question
benchmark containing 1,172 test questions. We evaluate accuracy by checking whether the correct
answer text or answer key letter appears in the model prediction.

\subsection{Baselines}

We compare our CRAG reproduction against a vanilla RAG baseline that uses the top-1 retrieved
document directly without any scoring or correction. For PopQA, retrieved documents come from the
Contriever retrieval results provided by the original CRAG repository. For ARC-Challenge, we
retrieve Wikipedia documents using our multi-stage Wikipedia search pipeline.

\subsection{Main Results}

Table~\ref{tab:main_results} presents our main results on both datasets.

\begin{table}[H]
\centering
\begin{tabular}{lcc}
\toprule
\textbf{Method} & \textbf{PopQA} & \textbf{ARC-Challenge} \\
\midrule
Vanilla RAG                  & 51.4 & 84.8 \\
CRAG (ours)                  & 54.4 & 85.2 \\
CRAG + Wikipedia (ours)      & 53.2 & ---  \\
\midrule
CRAG (original, LLaMA-2)     & 54.9 & 53.7 \\
\bottomrule
\end{tabular}
\caption{Accuracy (\%) on PopQA and ARC-Challenge. Our reproduction uses Phi-3-mini as the
generator. Original results are from \cite{yan2024corrective} using LLaMA-2-hf-7b. The large ARC
gap between CRAG (85.2\%) and vanilla RAG (84.8\%) is consistent with Phi-3-mini's strong
parametric science knowledge. The similar PopQA results across generators (54.4\% vs.\ 54.9\%)
suggest CRAG's correction mechanism is the primary driver of performance rather than
generator-specific capabilities.}
\label{tab:main_results}
\end{table}

Our open-source CRAG reproduction achieves 54.4\% on PopQA, closely matching the original
system's 54.9\% despite using a different generator. On ARC-Challenge, our system achieves 85.2\%
versus 84.8\% for vanilla RAG, a modest but consistent improvement.

\subsection{Action Distribution Analysis}

Table~\ref{tab:action_results} shows accuracy broken down by triggered action on PopQA.

\begin{table}[H]
\centering
\begin{tabular}{lccc}
\toprule
\textbf{Action} & \textbf{Count} & \textbf{\%} & \textbf{Accuracy (\%)} \\
\midrule
\textsc{Correct}   & 754 & 54.4 & 78.1 \\
\textsc{Ambiguous} & 379 & 27.4 & 19.3 \\
\textsc{Incorrect} & 252 & 18.2 & 36.1 \\
\bottomrule
\end{tabular}
\caption{Action distribution and per-action accuracy on PopQA (n=1,385). The \% column shows what
fraction of questions triggered each action; Accuracy (\%) shows correctness within that action
subset.}
\label{tab:action_results}
\end{table}

The \textsc{Correct} action achieves 78.1\% accuracy, a 26.7 percentage point improvement over
vanilla RAG (51.4\%), demonstrating that the T5 evaluator effectively identifies high-quality
retrievals. The \textsc{Ambiguous} action achieves only 19.3\% without web search, confirming that
web search is essential for this action. Adding Wikipedia search improves \textsc{Ambiguous}
accuracy to 23.0\%, a 4.7 percentage point gain.

On ARC-Challenge, the T5 evaluator classifies 88.3\% of questions as \textsc{Ambiguous},
reflecting its training distribution bias toward biographical entity questions rather than science
questions. Despite this, CRAG maintains competitive performance due to Phi-3-mini's strong
parametric knowledge of science.

\clearpage
\section{Explainability Analysis}

To understand what the T5 retrieval evaluator has learned, we apply SHAP to analyze token-level
attributions across all three action types. We select 9 representative samples (3 per action type)
and compute SHAP values using the \texttt{PartitionExplainer} with a text masker. We note that
these findings are based on qualitative case studies; future work should validate patterns across
larger samples with aggregate statistics.

\subsection{SHAP Results}

\begin{figure}[p]
\centering
\includegraphics[width=0.80\textwidth]{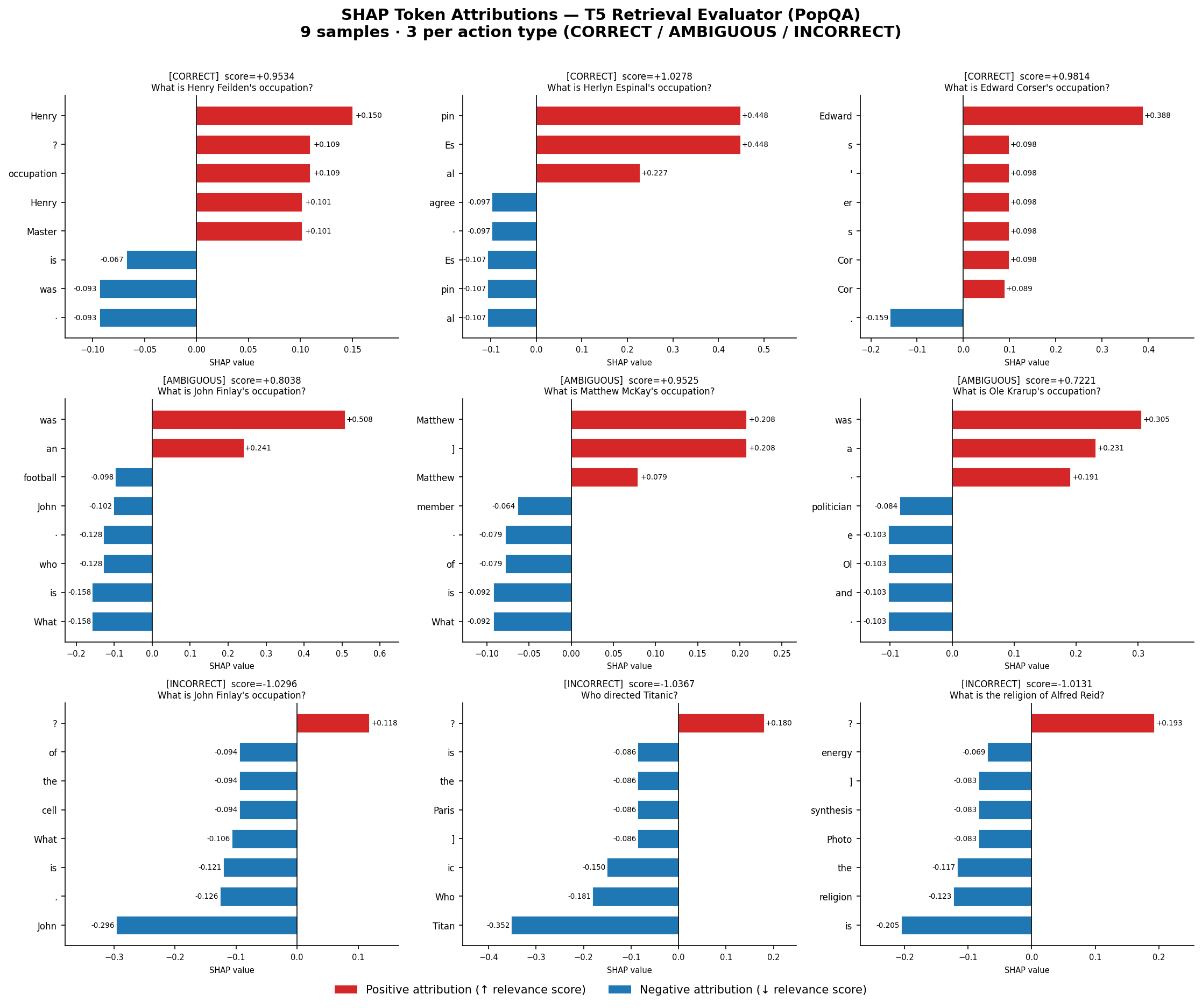}
\caption{SHAP token attributions for 9 samples (3 per action type). Red bars indicate positive
contributions to the relevance score; blue bars indicate negative contributions. Entity name tokens
dominate \textsc{Correct} samples positively and \textsc{Incorrect} samples negatively.}
\label{fig:shap}
\vspace{1em}
\includegraphics[width=0.90\textwidth]{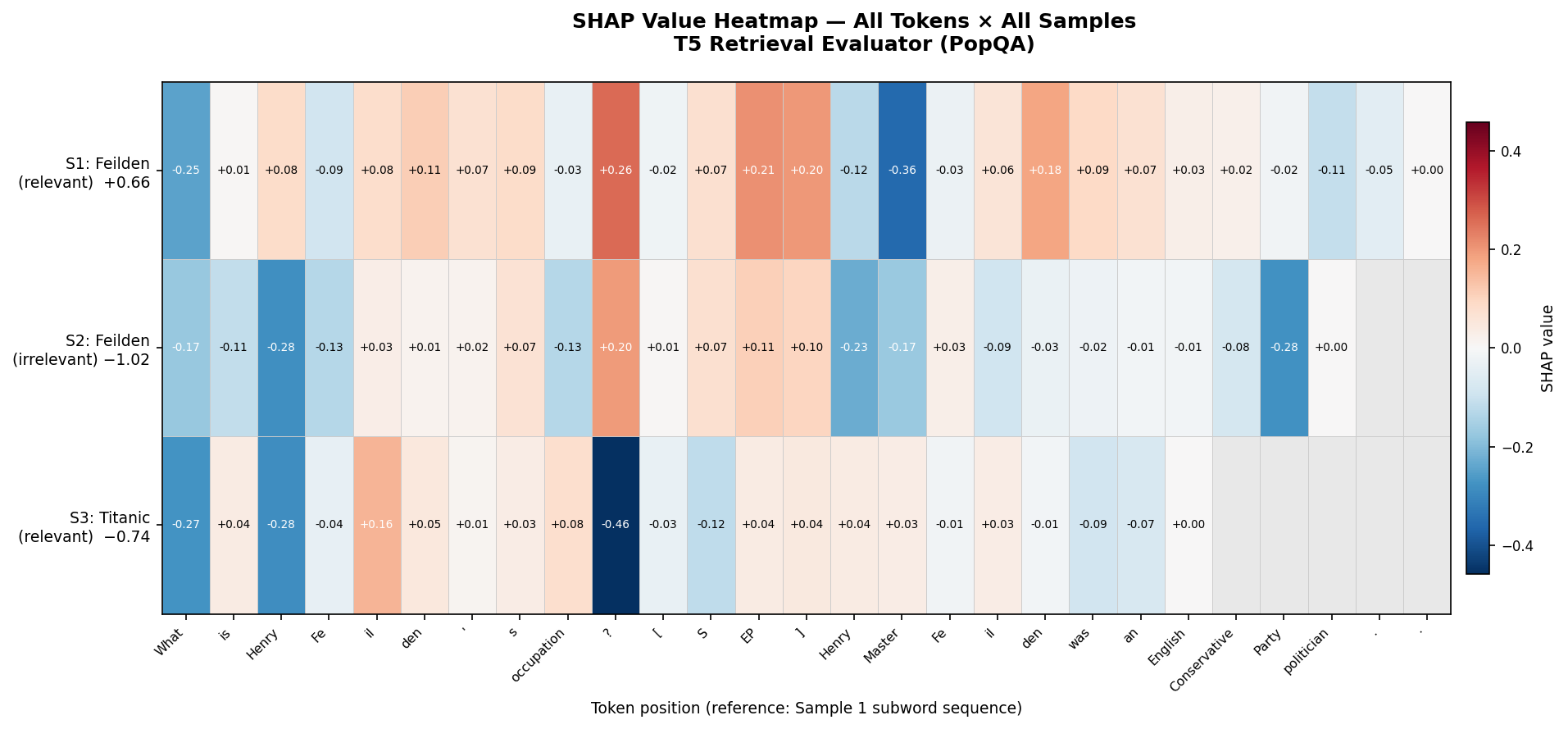}
\caption{SHAP value heatmap across token positions for three samples (\textsc{Correct},
\textsc{Incorrect}, \textsc{Ambiguous}). Red indicates positive attribution; blue indicates
negative. The question mark (\texttt{?}) token shows contrasting behavior between entity questions
(positive) and the Titanic question (strongly negative). Note: \texttt{·} symbols are SentencePiece
whitespace artifacts from the T5 tokenizer.}
\label{fig:shap_heatmap}
\end{figure}

Figure~\ref{fig:shap} shows token attributions for all 9 samples. Three consistent patterns emerge
across action types.

\textbf{Named entity alignment drives \textsc{Correct} scores.} For questions where the retrieved
document is relevant, entity name tokens contribute the largest positive SHAP values. For example,
for ``What is Henry Feilden's occupation?''\ paired with his Wikipedia article, the token
\textit{Henry} contributes $+0.150$ and \textit{occupation} contributes $+0.109$. This
demonstrates that the evaluator learned to match named entities between the question and document.

\textbf{Entity mismatch is the strongest \textsc{Incorrect} signal.} When a document is
irrelevant, the named entity token from the question becomes a strong negative driver. For the same
question paired with an irrelevant document about mitochondria, \textit{Henry} contributes
$-0.280$ and biology terms (\textit{mitochondria}, \textit{cell}) contribute negatively. The
evaluator detects irrelevance through entity absence rather than semantic content.

\textbf{The \texttt{[SEP]} separator may encode structural format.} The separator tokens
consistently receive positive SHAP values across \textsc{Correct} and \textsc{Ambiguous} samples.
This may suggest the evaluator learned the structural format of its training data, though this
pattern is also a common artifact in token-level SHAP on transformer models and warrants further
investigation.

\textbf{Out-of-distribution entities receive low scores.} For ``Who directed Titanic?''\ paired
with a factually correct document, the subword token \textit{Titan} (the first subword of
``Titanic'' in the question) receives $-0.459$, and the corresponding document subword receives
$-0.284$, consistent with the document's low relevance score, suggesting that movie-title entities
are underrepresented in the PopQA training distribution.

Figure~\ref{fig:shap_heatmap} shows a complementary heatmap view of SHAP values across all token
positions for the first three samples, revealing how attribution patterns shift across the
question-document boundary.

\subsection{Implications}

These findings reveal that the T5 evaluator functions primarily as an entity alignment detector
rather than a semantic relevance judge. This explains two observed failure modes: (1) factually
relevant documents about uncommon entity types such as movie titles and song names receive low
scores because their entity names are rare in the PopQA training distribution, and (2) the
evaluator struggles to transfer to science domains (ARC-Challenge) where questions lack named
person entities. Future work should fine-tune the evaluator on a more diverse set of question types
to address these limitations.

\section{Error Analysis}

\subsection{Performance by Question Type}

We categorize all 1,385 PopQA questions by type based on question keywords and analyze accuracy
per category. Figure~\ref{fig:error} shows accuracy broken down by question type, colored by the
dominant action triggered, and Figure~\ref{fig:heatmap} shows a detailed breakdown by action.

\subsection{Findings}

Table~\ref{tab:error} summarizes accuracy by question type.

\begin{table}[H]
\centering
\begin{tabular}{lccl}
\toprule
\textbf{Question Type} & \textbf{Count} & \textbf{Accuracy} & \textbf{Dominant Action} \\
\midrule
Country    & 274 & 85.8\% & \textsc{Correct}   \\
Sport      & 196 & 73.0\% & \textsc{Correct}   \\
Occupation & 121 & 61.2\% & \textsc{Correct}   \\
City       & 175 & 54.3\% & \textsc{Correct}   \\
Author     & 234 & 40.2\% & \textsc{Ambiguous} \\
Composer   &  46 & 32.6\% & \textsc{Ambiguous} \\
Director   &  90 & 28.9\% & \textsc{Ambiguous} \\
Genre      & 112 & 22.3\% & \textsc{Ambiguous} \\
Religion   &  40 &  5.0\% & \textsc{Correct}   \\
\bottomrule
\end{tabular}
\caption{CRAG accuracy by question type on PopQA (n=1,385). Dominant Action refers to the most
frequently triggered action for that question type.}
\label{tab:error}
\end{table}

Several patterns are notable. Question types where the T5 evaluator triggers \textsc{Correct}
action frequently achieve high accuracy --- country questions reach 85.8\% with 98\% accuracy on
the \textsc{Correct} action. In contrast, question types dominated by the \textsc{Ambiguous}
action (author, composer, director, genre) perform poorly at 22--40\%, confirming that these
categories require web search for reliable answers.

Religion questions are a special case: although the T5 evaluator most frequently assigns
\textsc{Correct} action (suggesting it finds at least one retrieved document above the confidence
threshold), overall accuracy is only 5.0\% across all action types. This indicates that high
evaluator confidence does not guarantee a correct answer for religion questions --- the retrieved
passages may be topically relevant but do not contain the specific answer entity, making answer
extraction unreliable.

\begin{figure}[p]
\centering
\includegraphics[width=0.88\textwidth]{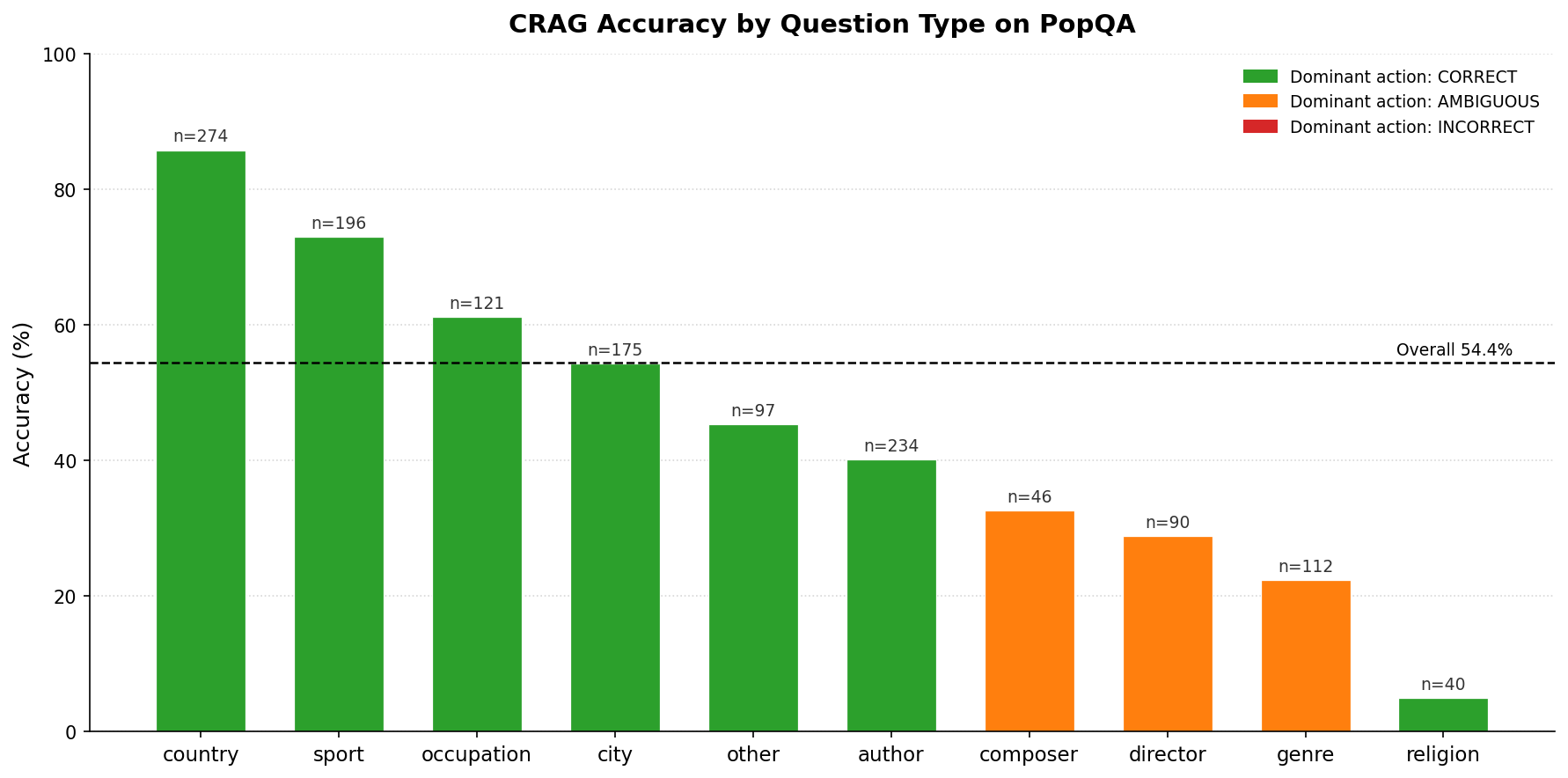}
\caption{CRAG accuracy by question type on PopQA. Green bars indicate \textsc{Correct}-dominant
question types; orange bars indicate \textsc{Ambiguous}-dominant types. The dashed line shows
overall accuracy (54.4\%).}
\label{fig:error}
\vspace{1em}
\includegraphics[width=0.72\textwidth]{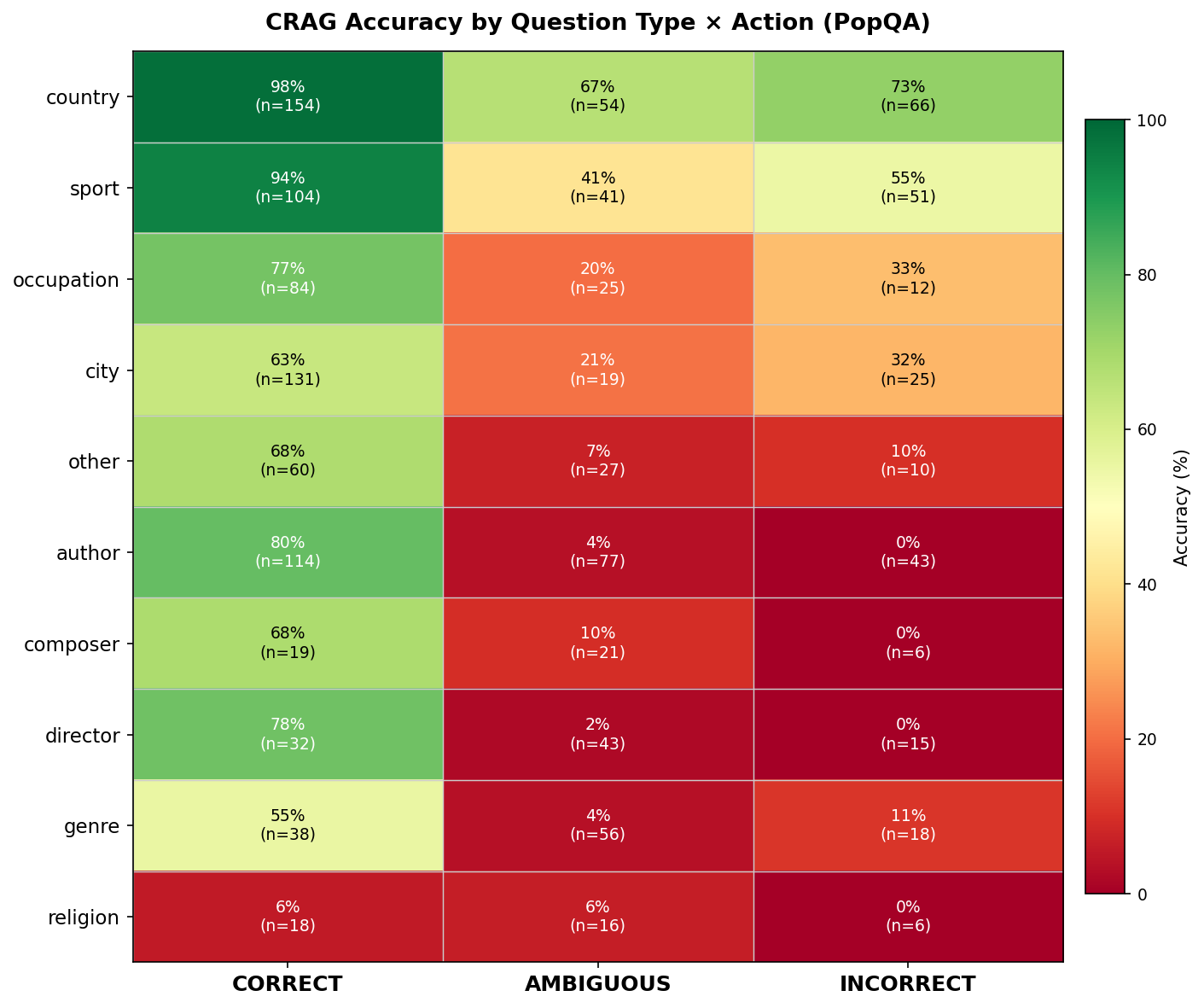}
\caption{Accuracy heatmap by question type and action. Each cell shows accuracy and sample count.
Green indicates high accuracy; red indicates low accuracy. The \textsc{Incorrect} action achieves
0\% for author, composer, and director question types.}
\label{fig:heatmap}
\end{figure}

The heatmap in Figure~\ref{fig:heatmap} reveals that the \textsc{Incorrect} action achieves 0\%
accuracy for author, composer, and director questions, suggesting that when the T5 evaluator
rejects all retrieved documents for these types, Phi-3-mini's parametric knowledge alone is
insufficient. Notably, the \textsc{Incorrect} action still performs reasonably well for country
(73\%) and sport (55\%) questions, where Phi-3-mini retains strong parametric knowledge even
without retrieved context.

\clearpage
\section{Limitations}

This work has several limitations that should be considered when interpreting our results.

\textbf{Single-run evaluation.} All results are based on a single experimental run without
confidence intervals. The 0.5\% gap between our PopQA result (54.4\%) and the original (54.9\%)
is within typical run-to-run variance and should not be interpreted as a meaningful difference.

\textbf{SHAP sample size.} Our explainability analysis is based on 9 qualitative case studies.
While the patterns are consistent across samples, larger-scale analysis with aggregate statistics
across 50--100 samples would provide stronger statistical support for our findings.

\textbf{No threshold tuning.} We use the original PopQA thresholds ($\tau^+ = 0.59$,
$\tau^- = -0.99$) without re-tuning for our generator or for ARC-Challenge. Domain-specific
threshold tuning may improve performance.

\textbf{Wikipedia vs.\ commercial web search.} Our Wikipedia API replacement achieves 82.3\% hit
rate on \textsc{Ambiguous} questions, leaving 17.7\% with no external context. A commercial web
search API would likely achieve higher coverage, potentially improving \textsc{Ambiguous} accuracy
beyond the 23.0\% we report.

\textbf{Generator comparison.} Our results use Phi-3-mini-4k-instruct as the generator, while the
original paper uses LLaMA-2-7B. Performance differences may reflect generator capability
differences rather than CRAG's correction mechanism. We mitigate this by reporting CRAG
vs.\ RAG deltas within the same generator.

\clearpage
\section{Conclusion}

In this work, we presented a fully open-source reproduction of CRAG, demonstrating that
proprietary components can be replaced with free alternatives while maintaining comparable
performance. Our reproduction achieves 54.4\% on PopQA versus the original system's 54.9\%, and
85.2\% versus 84.8\% for vanilla RAG on ARC-Challenge. We also built and released a five-stage
Wikipedia retrieval pipeline achieving 99\% document coverage on ARC-Challenge and 82.3\% on PopQA
\textsc{Ambiguous} questions --- a fully free replacement for the Google Search API used in the
original system.

Beyond reproduction, we contributed the first explainability analysis of CRAG's T5 retrieval
evaluator using SHAP. Our analysis reveals that the evaluator functions primarily as a named entity
alignment detector rather than a semantic relevance judge. This finding explains two key failure
modes: poor performance on out-of-distribution entity types such as movie titles and song names,
and systematic domain transfer failure on science questions, where 88.3\% of ARC-Challenge
questions are classified as \textsc{Ambiguous} because the evaluator was trained exclusively on
biographical entity questions.

Our error analysis across nine question types shows that \textsc{Correct}-dominant categories
(country, sport, occupation) benefit most from CRAG's correction mechanism, while
\textsc{Ambiguous}-dominant categories (author, director, genre) require web search to perform
well. Religion questions represent a systematic failure mode at 5.0\% accuracy despite the
evaluator frequently assigning \textsc{Correct} action, revealing that high retrieval confidence
does not guarantee answer correctness for non-entity answer types.

Future work should focus on three directions. First, fine-tuning the retrieval evaluator on more
diverse question types to improve domain transfer beyond biographical entity questions. Second,
investigating lightweight alternatives to commercial web search APIs that provide higher coverage
than Wikipedia alone for the \textsc{Incorrect} and \textsc{Ambiguous} actions. Third, conducting
larger-scale SHAP analysis with statistical validation across 50 or more samples per action type to
move beyond qualitative case studies.


\end{document}